# Development of sensitive long-wave infrared detector arrays for passively cooled space missions

Craig McMurtry
Donald Lee
James Beletic
Chi-Yi A. Chen
Richard T. Demers
Meghan Dorn
Dennis Edwall
Candice Bacon Fazar
William J. Forrest
Fengchuan Liu
Amanda K. Mainzer
Judith L. Pipher
Aristo Yulius

# Development of sensitive long-wave infrared detector arrays for passively cooled space missions

**Craig McMurtry**
University of Rochester
Department of Physics and Astronomy
Rochester, New York 14627
E-mail: craig@pas.rochester.edu

**Donald Lee**
**James Beletic**
**Chi-Yi A. Chen**
**Richard T. Demers**
Teledyne Imaging Sensors
5212 Verdugo Way
Camarillo, California 90312

**Meghan Dorn**
University of Rochester
Department of Physics and Astronomy
Rochester, New York 14627

**Dennis Edwall**
Teledyne Imaging Sensors
5212 Verdugo Way
Camarillo, California 90312

**Candice Bacon Fazar**
University of Rochester
Department of Physics and Astronomy
Rochester, New York 14627
  and
Roberts Wesleyan College
Department of Computer Science, Mathematics
  and Physics
2301 Westside Drive
Rochester, New York 14624

**William J. Forrest**
University of Rochester
Department of Physics and Astronomy
Rochester, New York 14627

**Fengchuan Liu**
**Amanda K. Mainzer**
California Institute of Technology
Jet Propulsion Laboratory, M/S 264-723
4800 Oak Grove Drive
Pasadena, California 91109

**Judith L. Pipher**
University of Rochester
Department of Physics and Astronomy
Rochester, New York 14627

**Aristo Yulius**
Teledyne Imaging Sensors
5212 Verdugo Way
Camarillo, California 90312

**Abstract.** The near-earth object camera (NEOCam) is a proposed infrared space mission designed to discover and characterize most of the potentially hazardous asteroids larger than 140 m in diameter that orbit near the Earth. NASA has funded technology development for NEOCam, including the development of long wavelength infrared detector arrays that will have excellent zodiacal background emission-limited performance at passively cooled focal plane temperatures. Teledyne Imaging Sensors has developed and delivered for test at the University of Rochester the first set of approximately 10 $\mu$m cutoff, $1024 \times 1024$ pixel HgCdTe detector arrays. Measurements of these arrays show the development to be extremely promising: noise, dark current, quantum efficiency, and well depth goals have been met by this technology at focal plane temperatures of 35 to 40 K, readily attainable with passive cooling. The next set of arrays to be developed will address changes suggested by the first set of deliverables. © *The Authors. Published by SPIE under a Creative Commons Attribution 3.0 Unported License. Distribution or reproduction of this work in whole or in part requires full attribution of the original publication, including its DOI.* [DOI: 10.1117/1.OE.52.9.091804]

## 1 Introduction: Requirements on Long-Wavelength Detector Arrays

The continued advancement of space-based astronomy/planetary science depends critically on the development of improved detector array technology. For example, the recently launched wide-field infrared survey explorer (WISE)[1] mission achieved point-source sensitivities that were improved by hundreds of times over the infrared astronomical satellite (IRAS)[2] which launched in 1983, despite the fact that the IRAS telescope's primary mirror was larger than WISE's (60 cm versus 40 cm). Spatial resolution improved at 12 $\mu$m from ∼0.5 arcmin for IRAS to ∼6 arcsec for WISE. The major factor responsible for the advances in both sensitivity and spatial resolution was the vast improvement in detector array technology, from 62 pixels for IRAS to WISE's 4 million pixels. Many astronomical and planetary applications will benefit from the construction of long-wave detector arrays with less demanding temperature requirements. One such example is the near-earth object camera (NEOCam),[3] a NASA discovery-class mission proposal designed to detect, discover, and characterize a large fraction of the asteroids and comets that most closely approach the Earth, the near-earth objects (NEOs).[4,5] NEOCam is a 0.5 m space telescope with a single imaging instrument operating at two wavelength ranges: 4 to 5 $\mu$m and 6 to 10 $\mu$m. Mid-wave 5 $\mu$m HgCdTe detector arrays for the James Webb space telescope (JWST) are designed to perform well at 37 to 46 K and require no further development for NEOCam. Our goal for the NEOCam mission has been to develop long-wave 10 + $\mu$m cutoff arrays that function at 35 to 40 K focal plane temperatures, with a preference for ∼40 K, for ease of thermal design. Clearly, this long-wave infrared (LWIR) array development has far greater future application than just NEOCam. Other long-wave arrays used for astronomy (e.g., Si:As) require focal plane temperatures ≤8 K (e.g., Spitzer IRAC/IRS/MIPS, WISE, Akari).

The NEOCam design allows the mission to operate throughout its four-year survey phase and beyond without the need for life-limiting cryogens or costly and potentially unreliable cryo-coolers. The Spitzer space telescope's focal planes[6–8] have equilibrated to ∼27 K now that its liquid helium cryogen supply has been exhausted, demonstrating the efficacy of passive cooling for a thermally well-designed space telescope. Throughout this warm mission Spitzer's two midwave InSb channels at 3.6 and 4.5 $\mu$m have performed with undiminished sensitivity, but require focal plane temperatures not much higher than 30 K. The WISE mission's 5.4 $\mu$m cutoff HgCdTe arrays, manufactured by Teledyne Imaging Systems (TIS), operated at 32 K with dark currents <5 e$^-$/s/pixel.[9]

In 2003, in response to a NASA grant awarded to University of Rochester (UR), TIS delivered prototype 512 × 512 pixel HgCdTe arrays operating out to 10 + $\mu$m. They were shown to have low dark current and low read noise for a moderately large fraction of the pixels.[10–13] For NEOCam, the fabrication of a mosaic of 1024 × 1024 HgCdTe arrays operating out to 10 $\mu$m is required to enable its wide-field survey. TIS's HAWAII-1RG (H1RG) readout integrated circuit (ROIC) technology was selected for NEOCam based on the readout family's heritage in space- and ground-based astronomical applications, including WISE,[14] the Hubble space telescope's wide-field camera-3,[15] and the JWST.[16,17] The HAWAII arrays have demonstrated the low power dissipation needed to support passive cooling as well as the low noise performance required to detect faint, natural background-limited astronomical sources and the narrow range of actual detector biases needed for operation of longer-wave HgCdTe photodiodes with a source-follower per detector ROIC.

The requirements for the NEOCam 10 $\mu$m arrays and comparison values for measurements of the three arrays tested are summarized in Table 1. Of course, detailed exposition of those measurements are found later in the text. The NEOCam telescope orbits the earth-sun L1 Lagrange point and can therefore achieve its required temperatures via passive cooling, so the dominant background signal in this channel is due to thermal emission from our solar system's zodiacal dust cloud. For observations on the solar system's ecliptic plane, the contribution from the zodiacal background increases with decreasing solar elongation. The minimum estimated background current for the 10 $\mu$m channel is ∼300 e$^-$/s/pixel therefore, dark current will not be a significant contribution to overall system noise if it is <200 e$^-$/s/pixel. Other missions will, of course, have different requirements.

In 2010, the NASA Discovery program awarded the NEOCam project technology funding to support the development of megapixel HgCdTe arrays operating with low noise and low dark current at wavelengths out to ∼10 $\mu$m or beyond at focal plane temperatures that could be achieved with passive cooling. The goals of the technology development program were fourfold: (1) to increase the array format from 512 × 512 to 1024 × 1024 pixels; (2) to deliver arrays with a cutoff wavelength of 10 + $\mu$m with excellent operability; (3) to meet well depth requirements; and (4) to maintain the low dark current and read noise characteristics of shorter wavelength TIS HgCdTe arrays. Meanwhile, NASA's APRA program funded the testing, characterization, and optimization efforts of the UR team. We report in this paper the outstanding results from the first phase of the LWIR detector development program.

## 2 Long-Wave Detector Array Development

In order to meet NEOCam's relatively low background long wavelength detector array requirements, JPL and the UR have worked with TIS to build upon our earlier progress in the development of low background 8.6, 9.2, and 10.3 $\mu$m cutoff HgCdTe detector arrays. That earlier work included characterization of high dark current pixels associated with various types of defects, and TIS has developed techniques to overcome these to a large extent. In the last decade, TIS has demonstrated extremely sensitive 5 $\mu$m cutoff HgCdTe detector arrays for WISE and JWST,[16,17] and many of the lessons learned in that development were also applicable to long wavelength HgCdTe array development. TIS grew HgCdTe wafers for the NEOCam detector array development program with two variations. Process evaluation chip (PEC) testing showed three excellent wafers, and one 1024 × 1024 pixel detector array from each wafer was hybridized to an H1RG ROIC. The three arrays were delivered to the UR for test in July 2012.

**Table 1** Minimum NEOCam requirements compared against measured properties of three arrays.

|  | NEOCam requirement | H1RG-16885 | H1RG-16886 | H1RG-16887 |
| --- | --- | --- | --- | --- |
| Detector material | HgCdTe; CdZnTe substrate removed | HgCdTe: Substrate not removed | HgCdTe: Substrate not removed | HgCdTe: Substrate not removed |
| Array format | 1024 × 1024 | By design | By design | By design |
| Cutoff wavelength ($\mu$m) | 10 | 10.6 | 9.9 | 9.9 |
| Operating temperature (K) | 35 to 40 | 35 | 35 | 35 |
| Dark current ($e^-$/s/pixel) | <200 | <200 | <200 | <200 |
| CDS read noise ($e^-$) | 50 | 22 | 22 | 22 |
| Responsive quantum efficiency (RQE) (%) | >60 | 65 non-AR coated, process evaluation chip (PEC) | 62 non-AR coated, PEC | 63 non-AR coated, PEC |
| Well depth ($e^-$) | >45 k | >55 k | >66 k | >75 k |
| Pixel operability (%) | >90 | 95[a] | 94[a] | 95[a] |

[a]Pixel operability measured does not include RQE, but we estimate that spec. will easily be satisfied once arrays are AR coated. Also note that ~90% of the pixels show dark currents <1 $e^-$/s/pixel.

## 2.1 Prior Development

$Hg_{1-x}Cd_xTe$ is a ternary compound, and the composition parameter $x$ can be varied to give the required long-wavelength cut-off. Shorter wave cut-off space astronomy devices are easier to fabricate,[18] and astronomy caliber HgCdTe detector arrays for low background applications have not been demonstrated beyond 5.4 $\mu$m, except for the progress shown in our 2003 development program. The first generation prototype 512 × 512 × 36 $\mu$m detector arrays from that program that were bonded to the H1RG ROIC[10–12] showed tremendous promise over the smaller format arrays developed in 2001[12,19]; however, they could not fully meet the specifications for NEOCam and other space astronomy missions. TIS had in place a well-developed long-wavelength program to produce arrays for high-background applications, but the low-background long-wavelength development was not as advanced. The circa 2003 devices delivered to the UR could only support very small reverse biases, severely limiting well depth, even though the pixel nodal capacitance was quite large for these relatively large pixels (~100 fF). Capacitive transimpedance amplifier (CTIA) ROICs are often employed for long-wave HgCdTe; they are subject to glow and higher noise because they are constantly powered up. With a CTIA ROIC, a small detector bias could be tolerated and thus maintain low dark current while maintaining a large charge capacity. However, we rejected this ROIC choice, partially because of high CTIA power dissipation (excessive power dissipation for NEOCam requirements), and partially because there is unavoidable Johnson noise from low resistance pixels.

Since 2003, Teledyne has made significant strides in improving the operability of its LWIR detectors operating under low backgrounds and temperatures through true bandgap engineering, with independent control of both compositional profiling and doping to achieve both optimal electrical and optical performance. This control exhibits a number of advantages for LWIR low-background space missions. The wide bandgap cap reduces the susceptibility to surface-induced excess currents associated with variation in quality of passivation. This helps to improve the reproducibility of the process as well as to minimize the total detector dark current at low temperatures. Recent innovations by TIS in its MBE growth process have led to an extremely low density of defects responsible for tunneling currents, the primary source of excess detector current in reverse bias and the root cause for inoperability at these wavelengths, backgrounds, and operating temperatures. The 2012 LWIR detector arrays delivered by TIS also benefited from process improvements that were made after detector degradation was noticed by the JWST program.[20] The failure mode has been identified by TIS and has been fully eliminated through process modifications. The improved approach has already been implemented in most Teledyne FPA products for several years, with no reports of similar failures. This new approach is presently being used on all new JWST flight parts, as well as for NEOCam FPAs reported in this publication. Additionally, the NEOCam FPAs employ a proprietary architecture that has thus far completely eliminated this failure mode.

In Table 2 we list the dark currents and well depths achieved from the 2001 and 2003 developments, as well as the spectacular performance of the 2012-produced arrays we report on here (in bold). In this table we have emphasized dark current performance according to NEOCam requirements. In discussion on the individual arrays it will be seen that far better dark currents at slightly reduced operability are achieved.

UR utilizes a helium test dewar (see Fig. 3 of Forrest et al.[21]) with a ZnSe window, a filter wheel containing cold (~4 K) circular variable interference filters (4 to 8 and 8 to 14.3 $\mu$m, with bandwidths of 1.1% and 1.7%, respectively), several discrete filters, and a cold dark slide for dark current and noise measurements. A Lyot stop of 67.6 $\mu$m diameter defines the cold (∼4 K) aperture stop, and it is located 43.8 mm from the front surface of the array. The array is heated to temperatures of 30 to 45 K as monitored by a Lake Shore Cryotonics temperature controller. A 77 K can surrounds the helium can. We utilize an array controller, based on open-source hardware from the Observatories of the Carnegie Institute of Washington design, described by Moore et al.[22] Quantum efficiency measurements are made through the dewar window viewing a high emissivity black felt cloth at known temperature ∼300 K located below the dewar window. A measurement of the black cloth cooled to 77 K is subtracted from the ∼300 K measurement to isolate the in-band radiation. See Appendix for further description and validation of the technique.

Table 2  Dark currents and well depths for 10 μm arrays.

| Format | Pixel pitch (μm) | $\lambda_{cutoff}$ (μm) | Year produced | T | Well depth (e⁻) | Dark current (e⁻/s) | Operability (%)[a] | References |
|---|---|---|---|---|---|---|---|---|
| 256² | 40 | 9.3 | 2001 | 30 K | >49 k | <100 | 30[b] | Bacon et al.[19] |
| 512² | 36 | 9.3 | 2003 | 30 K | >25 k | <30 | 75 | Bacon et al.[11] |
|  |  |  |  |  | >50 k | <30 | 51 |  |
| 512² |  | 10.3 | 2003 | 30 K | >42 k | <100 | 70 | Bacon et al.[12] |
|  |  |  |  |  | >84 k | <100 | 29 |  |
| H1RG-16885 1024² | 18 | 10.6 | 2012 | 30 K (35 K) | >42 k | <200 | 97 (98) | This paper |
|  |  |  |  | 30 K (35 K) | >55 k | <200 | 95 (95) |  |
|  |  |  |  | 42 K | >50 k | <200 | 94 |  |
| H1RG-16886 1024² | 18 | 9.9 | 2012 | 35 K | >39 k | <200 | 99 | This paper |
|  |  |  |  | 35 K | >66 k | <200 | 94 |  |
| H1RG-16887 1024² | 18 | 9.9 | 2012 | 35 K (40 K) | >47 k | <200 | 99 (98) | This paper |
|  |  |  |  | 35 K (40 K) | >75 k | <200 | 95 (92) |  |

[a]Operability is based only on dark current for the well depth lower limit quoted. However, we fully expect future AR-coated arrays to meet quantum efficiency specification as well.
[b]This array was a test array with strips, comprised of differing diode geometries. The quoted result is only for the best strip (and this result informed future developments).

### 2.2 H1RG-16885

H1RG-16885 is a 10.6 μm cutoff array, constructed from one of the layers grown by Teledyne, which exhibited excellent Hall and defect characteristics and which was measured to have bulk doping density close to the target value as determined from the correlation of tunneling dark currents with doping density.[23] The array was bump-bonded to an H1RG ROIC, utilizing proprietary bonding to avoid defects at the diode implant,[24] and with a design introducing extra capacitance above that of the diode. For this first set of deliveries, the array was not epoxy backfilled and not substrate removed.

Nodal capacitance per pixel is required to calibrate quantum efficiency, well depth, dark current, and noise data, and is determined using the Signal versus $\sigma^2$ method[25] and shown in Fig. 1 for H1RG-16885. Averages for 70 regions of 50 × 50 pixel boxes are evaluated, and a typical set plotted.

The interpixel capacitance (IPC) correction for H1RG-16885 is determined using very high dark current pixels and the autocorrelation method.[26] It is important to note that this device is not epoxy backfilled: for use in space, the devices must be epoxy backfilled so that substrate removal can take place. The epoxy backfill impacts the value for the IPC. For this array without epoxy backfill, the average nearest-neighbor coupling parameter $\alpha$ is ∼1.2% and with epoxy backfill, rises to $\alpha$ ∼1.6%% using the hot pixel method. Thus the determined nodal capacitance for H1RG-16885, which is not epoxy back-filled, must be reduced by the factor $(1 + 8\alpha)$, or ∼10%.[26] The epoxy will increase this factor to ∼13% with backfill. NEOCam arrays will be epoxy backfilled and substrate removed. When that occurs, the capacitance and IPC will be remeasured using several methods.[27] All of the measurements reported here were conducted on arrays that were not epoxy backfilled unless noted. However, when using the autocorrelation method, we obtain $\alpha$ ∼2.0% for the case without epoxy backfill. The autocorrelation method employs a uniform illumination at moderate integrated fluence levels, i.e., not dark backgrounds as in the hot pixel method. The discrepancy (1.2% versus 2.0%) that we see between the two methods is consistent with the findings of other authors.[28–31]

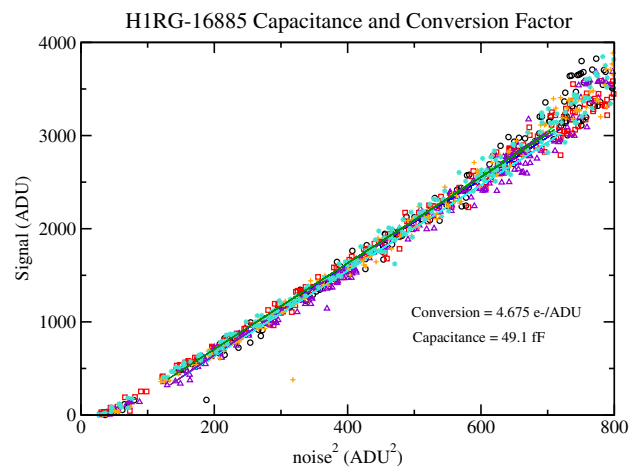

**Fig. 1** Capacitance per pixel determined for H1RG-16885, with no IPC (interpixel capacitance) correction applied.

PEC data from TIS gave a responsive quantum efficiency (RQE) of 65% at 100 mV of reverse bias and in the range from ∼4 to 9 $\mu$m, and in addition provided a cutoff wavelength of 10.6 $\mu$m at 30 K. It is important to note that these first three arrays did not include antireflection coating, thus 65% is very good (the theoretical maximum value possible is ∼75%). The UR has confirmed both the RQE (at 8 to 9 $\mu$m) and the half power cutoff wavelength with photometry of a black felt cloth of known temperature, through a cold pinhole of known dimensions and distance from the array, through a cold circular variable interference filter and a warm dewar window (Appendix).

Well depth measurements are obtained by exposing the array to a low constant flux at 8.6 $\mu$m monitoring the photo-signal (in mV) as a function of time to saturation. We define the well depth as the signal above which the increase in signal per unit time is less than half its initial (i.e., low signal) value. It should be noted that the H1RG ROIC adds a pedestal injection of 0 to 40 mV to our applied reverse bias to give the actual reverse bias at the beginning of the integration. For small dark currents, the pixel well depth equals the actual reverse bias for that pixel. Dark current is, of course, the most challenging requirement for space astronomy missions. One desires a dark current per pixel less than the zodiacal emission signal per pixel. NEOCam's dark current requirements are not nearly as stringent as are those of JWST at 5 $\mu$m, where NIRSpec requires <0.01 $e^-$/s/pixel to be background emission limited. NEOCam, because of the broad bandwidth of its long-wave channel, 6 to 10 $\mu$m (10 $\mu$m is near the zodiacal emission maximum) requires only that dark current be <200 $e^-$/s/pixel. The detector arrays we report on here easily meet NEOCam dark current operability requirements at 200 $e^-$/s/pixel and, in fact, meet much more stringent requirements at marginally reduced operability. We have measured the dark current and well depth of these devices at focal plane temperatures of 30 to 47 K, and for a variety of applied reverse biases. These are the first low background 10 $\mu$m HgCdTe arrays we have encountered whose diodes can withstand substantial reverse biases. In Figs. 2 and 3 we give examples of the dark current data for H1RG-16685. In Fig. 2, at an applied reverse bias of 200 mV, we provide a plot of well depth versus dark current for a focal plane temperature of 30 K. This atypical graph is our diagnostic based on the properties of the 2003-era deliveries where many very low dark current pixels turned out to also have nearly zero well depth. The concentration of points centered at 0.1 $e^-$/s/pixel and 240 mV well depth was our first indication that these 2012-era arrays did not have that problem. A well depth of 160 mV corresponds to ∼45,000 $e^-$: most of the array pixels are concentrated at a well depth exceeding the 200 mV applied bias (because of the H1RG ROIC pedestal injection of 0 to 40 mV), and with very low dark current. Some pixels have dark currents up to ∼100 $e^-$/s/pixel at that same well depth range as the concentration of pixels. In addition, there are a small percentage of pixels with dark currents of >100 $e^-$/s/pixel, which have debiased in the approximately 5 s between pixel reset and the first pedestal sample, reducing their well depths. Figure 3 gives the cumulative histogram of the dark current for well depths exceeding 180 mV and at a focal plane temperature of 35 K (vertical axis gives the percentage of 1016 × 1016 infrared active pixels). As can be seen, over 92% of the pixels have dark currents ≤1 $e^-$/s/pixel with a well depth >180 mV (>49,500 $e^-$), a truly spectacular achievement. At the NEOCam requirement of <200 $e^-$/s, the operability is 95%. Table 2 provides comparative data on the dark current operability at 30, 35, and 42 K for this array. The two other arrays showed similar slight degradation with increasing temperature up to 42 K.

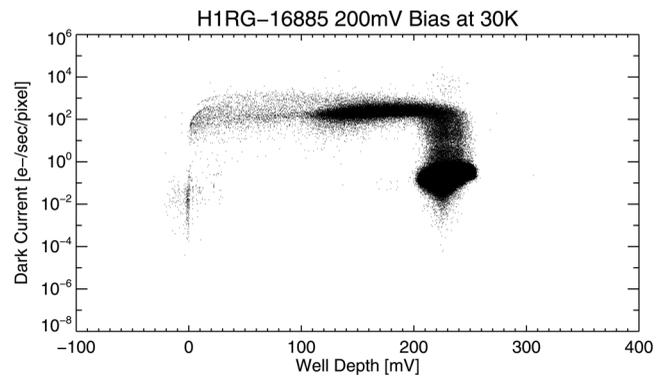

**Fig. 2** Well depth versus dark current for H1RG-16885 at a focal plane temperature of 30 K, and 200 mV applied reverse bias. As noted in text, this graph is a diagnostic, illustrating that the preponderance of pixels have excellent well depth and very low dark current and are centered around 0.1 $e^-$/s and 230 mV. Even for slightly higher dark current, there is only a small percentage of pixels that decrease in well depth because of debias between the pixel reset and the pedestal (first sample) measurement.

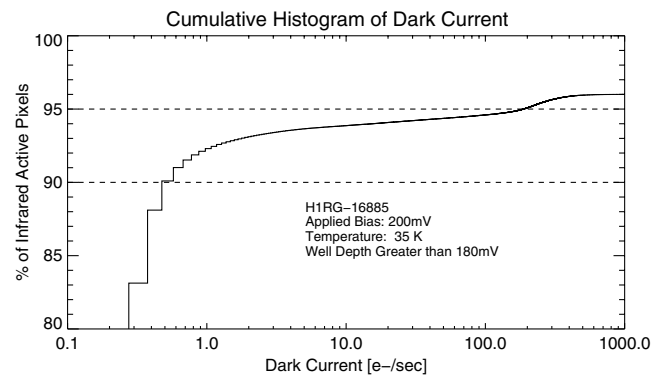

**Fig. 3** Histogram of the cumulative dark current for H1RG-16885 for well depths >180 mV for 200 mV reverse bias at a focal plane temperature of 35 K. Only infrared active pixels are included.

Dark current operability versus temperature was determined for focal plane temperatures of 30 to 47 K under the NEOCam dark current requirements of <200 $e^-$/s/pixel and is presented in Fig. 4. The ∼10 $\mu$m arrays can be operated successfully to focal plane temperatures ∼42 K, a fact that will positively impact NEOCam's ultimate thermal design. In Fig. 5, we show an Arrhenius plot of dark current versus temperature for SCA H1RG-16885. This plot also illustrates the excellent low dark current performance even at moderately high temperatures for the present generation of long-wave HgCdTe detector arrays. These data are the dark currents for the best performing pixels, i.e., those pixels with the lowest dark current and 240 mV of actual reverse detector bias, which is also equivalent to 240 mV of well depth. The data represented by filled circles were taken at a stable, constant temperature, which allows us to accurately measure the lowest dark currents achieved. The data represented by squares were taken while the temperature was allowed to slowly increase, which allows us to accurately measure the higher levels of dark current at those higher temperatures. The lines plotted are the models that were fitted to our data. At temperatures above 42 K, diffusion dark current is dominant, while generation-recombination dark current is the majority component over the range of 36 to 42 K. Below 36 K, there is a contribution to the apparent dark current that does not vary with either temperature or applied detector bias but instead behaves as if it were due to actual photons, e.g., a light leak or a glow from the unit cells of the ROIC at the 0.16 e−/s/pixel level. We believe that the light leak explanation is most probable. We have not fully investigated the source of these photons since a dark current of 0.16 e−/s/pixel is a full three orders of magnitude better than the requirements for NEOCam. Further, since we saw no variation in dark current with applied detector bias, then any trap-to-band or band-to-band tunneling dark currents must be at least an order of magnitude lower, i.e., <0.01 e−/s/pixel at 30 K for the given 240 mV detector bias shown in Fig. 5. It is extremely gratifying that these pixels have a negligible tunneling component to the dark current, unlike earlier generations of LWIR devices.[19]

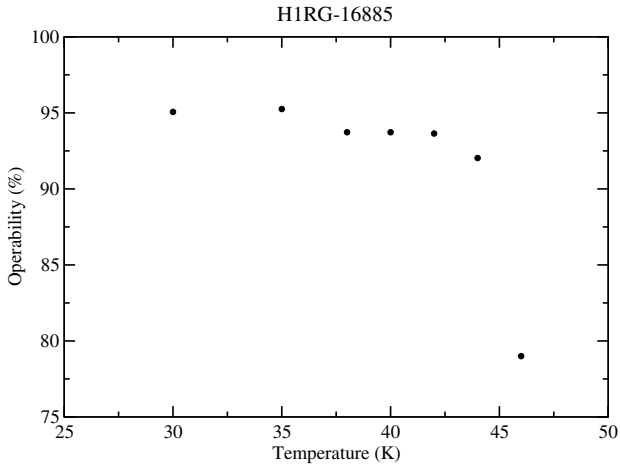

**Fig. 4** Operability versus focal plane temperature for H1RG-16885.

Overall pixel operability of the array includes not only dark current but also noise and quantum efficiency. While we did not calculate overall operability, we next describe measurements of the noise and RQE. Noise is determined from a data set consisting of 25 independent dark images of 16 sample-up-the-ramp frames. From this we create 25 Fowler-1, 25 Fowler-4, and 25 Fowler-8 images[25] from which the RMS read noise per pixel is calculated. Cosmic ray hits and random telegraph noise are removed through iterative $4\sigma$ clipping. Data were obtained at several focal plane temperatures. A histogram of 35 K Fowler-1 noise for H1RG-16885 is shown in Fig. 6; the median noise is 21.9 e− and reduces with increased number of samples. At 30 K focal plane temperature, the median Fowler-1 noise is slightly higher, 24.0 e−, probably because the H1RG was designed for higher temperature operation. The NEOCam noise requirement is <50 e−. All pixels with dark current <200 e−/s satisfy this and are operable on this count.

An image under flat field illumination through a cold Lyot stop (300 K black felt cloth) is shown in Fig. 7, corrected for two effects: the geometry and various corrections to produce a "flat field" image are discussed in Appendix A. A $\cos^4(\theta)$ drop-off in intensity from the optic axis has been removed, as well as vignetting attributed to the fact that our Lyot stop aperture of 67.6 μm has a thickness comparable in size to the diameter. The implied vignetting factor from the center (1.0) to the corner is 0.62 (see Appendix). A residual gradient in the image remains, however, and some remaining interference from the cold circular variable filter wheel centered at 5.6 μm. Implicit in our flat field and quantum efficiency estimates is the assumption that the black cloth is an effective blackbody. While this irradiance is not cavity radiation, our experience on Spitzer IRAC has shown that the assumption is valid: in-flight calibration against known standard stars yielded the expected response as measured in this manner in our lab.

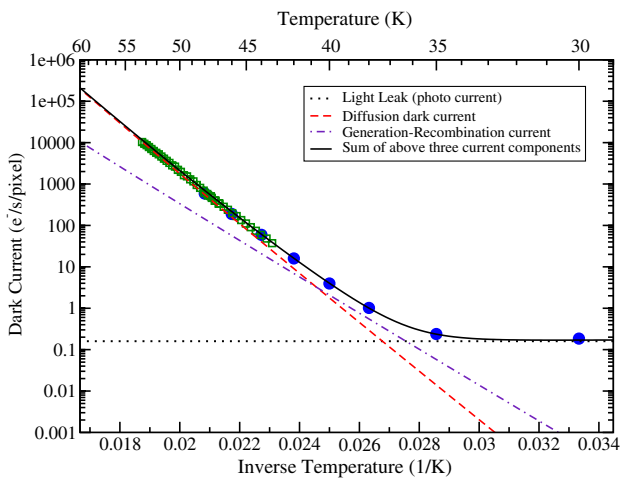

**Fig. 5** An Arrhenius plot of dark current per pixel versus inverse temperature for an actual well depth (reverse detector bias) of 240 mV.

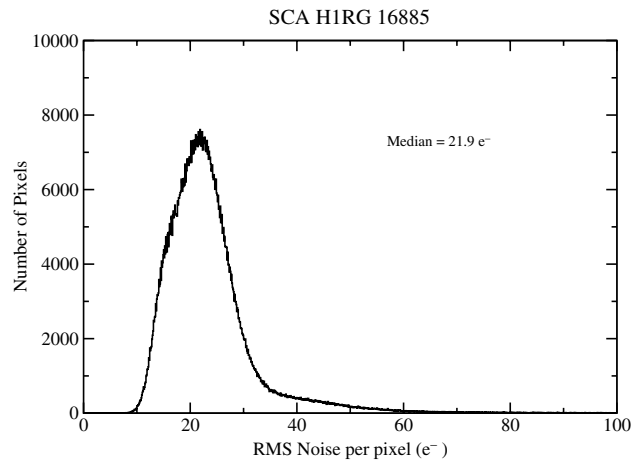

**Fig. 6** Histogram of Fowler-1 noise/pixel for H1RG-16885. Fowler-1 sampling is equivalent to correlated double sampling.

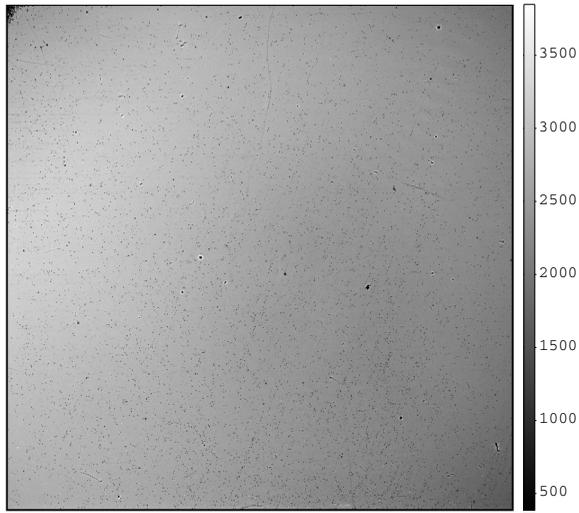

**Fig. 7** Flat-field illumination of H1RG-16885 through a circular variable filter wheel centered at 5.6 μm. The dark spots with bright halos are diffraction/shadows from dust specks suspended 800 μm in front of the detector surface, demonstrating the good imaging properties of the array.

RQE is calculated from these flat-field data. RQE as a function of wavelength (from 8 to 10 μm in the central 50 × 50 pixels[2] was found to verify within uncertainties both the RQE between 8 and 9 μm, and the cutoff wavelength as determined by TIS PEC results via a Fourier transform infrared (FTIR) spectrometer, for the wafer from which H1RG-16885 was constructed. We did not calculate total pixel operability including individual RQE/pixel because the device was not antireflection (AR)-coated, nor was the response perfectly "flat" due to illumination that was not properly uniform across the entire array. The NEOCam specification (for an AR-coated array) is >60%.

### 2.3 H1RG-16886

We obtained similar data on this device, an array with a 10.1 μm cutoff wavelength at 30 K, and RQE ~62% as measured by TIS from 4 to 9 μm. The responsive RQE value between 8 to 9 μm in our tests is ~58% (excellent agreement within uncertainties), but the cutoff wavelength appears to be ~9.9 μm at 35 K focal plane temperature (see Fig. 8) as compared with TIS determined 10.15 μm using an FTIR spectrometer at 30 K, which can be explained by the variation of the band gap energy with temperature.[28] The dark current operability for this array exceeds that of H1RG-16885 as can be seen from Table 2, and the cumulative dark current distribution for 35 K focal plane temperature, >180 mV well depth is shown in Fig. 9. Here the applied bias is 250 mV rather than 200 mV as displayed in Fig. 3 for H1RG-16885, and the performance is relatively degraded by this 25% higher bias. Following test, H1RG-16886 was sent back to TIS, and the area between the detector array material and the multiplexer was half back-filled with epoxy. The IPC nearest-neighbor coupling $\alpha$ was measured in the epoxied area using the hot pixel method and compared against that determined in the nonepoxied area. In Fig. 10 is a histogram showing the peak values determined for the two areas, with 1.2% corresponding to the nonepoxied area and 1.6% to the epoxied area.

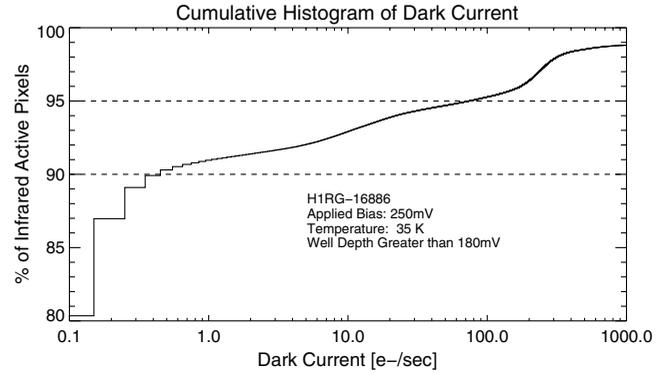

**Fig. 9** Histogram of the cumulative dark current for H1RG-16886 for well depths >180 mV for 250 mV reverse bias at a focal plane temperature of 35 K. Only infrared active pixels are included.

### 2.4 H1RG-16887

We obtained similar data on this device, an array with a 9.8 μm cutoff wavelength, and QE ~63% measured by TIS from 4 to 9 μm. As before, we roughly confirm these values in our tests. The dark current operability for this array is similar to those of H1RG-16885 and H1RG-16886 as can be seen from Table 2, and the cumulative dark current distribution for 35 K focal plane temperature, and >180 mV well depth is shown in Fig. 11. Again, the higher detector bias of 250 mV slightly degrades the dark current performance relative to that for the 200 mV bias employed for H1RG-16885.

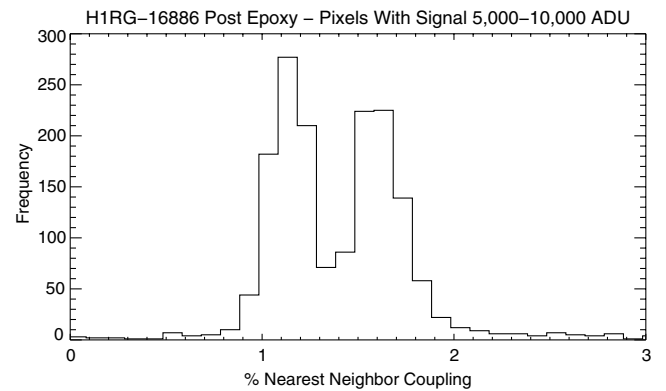

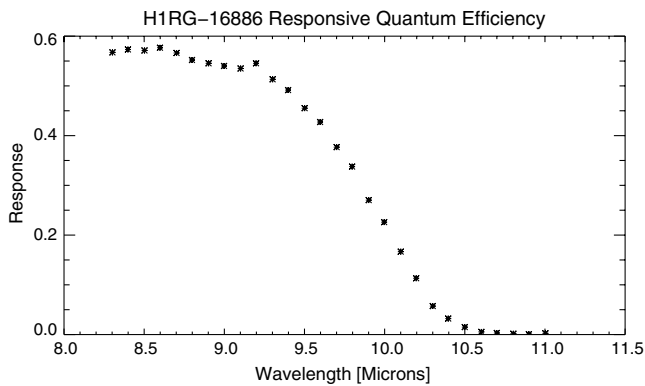

**Fig. 8** Responsive quantum efficiency (RQE) as a function of wavelength for H1RG-16886 at a focal plane temperature of 35 K.

**Fig. 10** Histogram of IPC nearest neighbor coupling parameter $\alpha$ obtained for H1RG-16776 for the half of the array that was not back-filled with epoxy (peak 1.2%) and for the half of the array that was (1.6%).

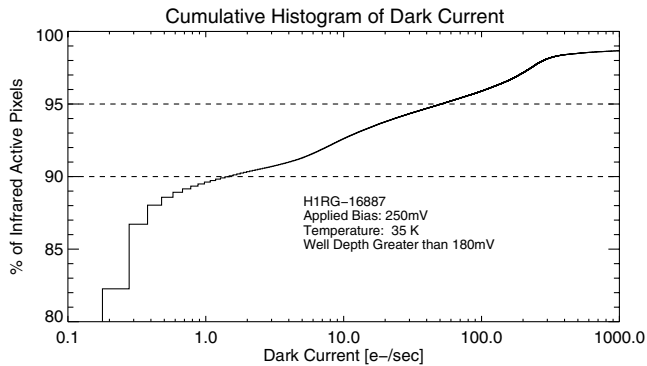

**Fig. 11** Histogram of the cumulative dark current for H1RG-16887 for well depths >180 mV for 250 mV reverse bias at a focal plane temperature of 35 K. Only infrared active pixels are included.

## 3 Summary

TIS has produced and delivered three superb low-background long-wave arrays to meet NEOCam specifications. Testing of the three arrays is reported on here with top-level results shown in Table 1. Future work will include enhanced IPC estimation, substrate removal, antireflection coating, and radiation testing.

*Acknowledgments*

A. Mainzer and the JPL group acknowledge support from NASA's Discovery program to fund development and acquisition of long-wave low-background arrays from Teledyne, and J. Pipher and the Rochester group acknowledge support by NASA APRA grant NNX12AF39G for test and optimization of those arrays.

## Appendix: Flat Field Measurement and Corrections

We use radiation flux from a room temperature blackbody to measure the quantum efficiency of the HgCdTe photodiode pixels. The blackbody is a black felt cloth with temperature measured to 0.1°C. To account for the cloth emissivity being less than 1.00, we rely on the temperatures of all the other objects in the room being at about the same temperature as the black cloth. The wavelengths of the incident photons are defined by a 4.2 K temperature interference filter approximately 45 mm from the detector's surface. The basic geometry of our setup is given in Fig. 3 of Forrest et al.[21] For lab measurements, the reimaging lens in that figure is removed and the black cloth is placed at the position of the "Image of Celestial Object." The technique of imaging black felt was utilized in our tests of Spitzer InSb detector arrays. Once these arrays were flown, and calibrated with measurements of several stars, the response in the IRAC bands centered at

3.6 and 4.5 $\mu$m, proved to be in good agreement with our lab estimates of the RQE. Furthermore, the TIS measured RQE of the PECs were within 6% of our measurements; the TIS measurements employed a NIST-calibrated source.

To limit the stray radiation in the cold "inner sanctum" housing the detector, we use a Lyot stop defined by a 67.6 $\mu$m diameter hole through a 50 $\mu$m thick metal foil. For pixels on the optic axis, the pixel area is taken to be that of an 18 $\mu$m square. The solid angle is given by the effective optical distance between the detector's back surface and the 67 $\mu$m Lyot stop. The physical distance is 44.6 mm. The 0.8 mm thick CdZnTe substrate acts to reduce that distance by 1.3 mm. At an angle $\theta$ off axis, the effective pixel area is reduced by the projection factor $\cos(\theta)$ while the effective solid angle is reduced by the factor $\cos^3(\theta)$. In addition, the finite thickness of the Lyot stop metal (50 $\mu$m) leads to vignetting of the full aperture, for off axis angles $\theta$, by the factor

$$\frac{2}{\pi}[\arccos(X) - X\sqrt{(1-X^2)}]$$

where

$$X = \frac{dx}{2r}$$

and $r$ is the radius of the Lyot stop, $dx = t\tan(\theta)$ is the apparent offset between circles at the top and bottom of the metal foil, $t$ is the thickness of the foil (50 $\mu$m), and $\theta$ is the angular deviation from the optic axis.

At the corners of the array, $\theta$ is approximately 16 deg. For pixels at the corners, the $\cos^4(\theta)$ reduction factor is 0.85 and the vignetting factor is 0.73, for a total loss in photons of 0.62 compared with those seen by the pixels on the optic axis.

To account for other photons (e.g., from the dewar window, or warm filters), we substitute the room temperature black cloth with a black cloth submerged in liquid nitrogen. The in-band photons from such a low temperature should be negligible compared with the room temperature load. This signal is subtracted from the measurement of the room temperature black cloth accounting for other sources of radiation inside the dewar.

Our systematic uncertainties for RQE measurements are approximately 10% (absolute), although the actual RQE value reported here is a conservative lower limit.

Biographies and photographs of the author are not available.